\documentstyle[epsf,11pt]{article}
\def\ll{\label}
\def\re{\ref}
\def\c{\cite}

\def\r1{(\ref{$1})}

\def\ti{\tilde}

\def\ba{\begin{array}{c}}

\def\ea{\end{array}}

\def\bet{\beta}
\def\ov{\over}
\def\ha{{1\over 2}}

\def\l{\left}
\def\l({\left(}
\def\r){\right)}
\def\r{\right}

\def\la{\lambda}
\def\al{\alpha}
\def\be{\begin{equation}}
\def\bc{\begin{center}}
\def\ec{\end{center}}
\def\ee{\end{equation}}
\def\ed{\end{document}}
\def\bea{\begin{eqnarray}}
 \def\eea{\end{eqnarray}}

\def\bi{\begin{itemize}}
 \def\ei{\end{itemize}}

\begin{document} 
\title{    
   Consistent refinement of 
Bethe strings for spin and electron models   and 
 a new non-Bethe solution  }
\author{
Anjan Kundu \footnote {email: anjan@tnp.saha.ernet.in} \\  
  Saha Institute of Nuclear Physics, \\
 1/AF Bidhan Nagar, Calcutta 700 064, India.
 }
\maketitle
%

\begin{abstract} 

 Well known Bethe strings in  spin chains, $\delta$-electron gas, Hubbard
and $t-J$ models are shown to be imprecise, while their consistent
 refinement  along with a  new non-Bethe $r$-string  are discovered.
Connection with the earlier results is established and   
 the string
 hypothesis problem is discussed in the light of the present findings .

\end{abstract}

\medskip

cond-mat/0108175

\medskip


A new direction in physics, namely the theory of
 exactly solvable quantum systems opened up  following the 
classic work of Bethe \c{bethe31}. Though 
 Bethe's objective was to  solve
exactly the eigenvalue problem of  Heisenberg spin chain only,
this $XXX$ spin-$\ha$ model proved to be a generic one and 
 almost the same Bethe ansatz or its nested extensions to include particles
with more degrees of freedom, e.g. electrons, were found to be applicable 
to a number of important models in condensed matter physics
as well as in  quantum field theory \c{review}.
The celebrated
Bethe ansatz
 assumes a specific form for the  eigenvectors having  parametric
dependence on rapidity     
 variables, which in turn are to 
 be determined   from specified equations known as 
the  Bethe ansatz equations (BAE).
Bethe  proposed also  a special type of
complex solutions for BAE at the  thermodynamic limit in the form of a
   string, known as the Bethe string. He  showed further that such strings
 form a complete set  by proving that their total number   
together with the  real roots 
give   exactly
 the required number of Bethe ansatz states.
These Bethe string solutions played important role in all the Bethe ansatz 
solvable   fundamental  integrable models like  spin chains
 \c{taka71a,gaudin71},
 $\delta$-function electron gas
\c{taka71b,lai71}, Hubbard model \c{taka72,kor94}, supersymmetric
$t-J$ model \c{schlot87,foers93} etc.
The string solutions correspond to the    bound states  and 
 describe excited states of
the models. 
A popular conjecture,  known as
the { string hypothesis}, assumes that 
all complex solutions of BAE at the thermodynamic limit must    be
 of the  Bethe string type. Though this conjecture was never really
proved, it was used successfully  to all the above models for calculating 
important physical quantities like energy, entropy, pressure,
magnetization etc.
 through thermodynamic Bethe ansatz (TBA). 
However  in early eighties  some criticisms started
appearing in this success story of the Bethe string,
 when in a series of studies
the string hypothesis apparently failed and instead of the Bethe strings
 non-Bethe complex solutions were reported to  appear as  close-roots in the
form of quartets and 2-strings and as wide-roots with no restriction on
their forms \c{DesLow82,Woy82,BabdeVVia83,Woy83}.
 In a subsequent paper  \c{vlad84} an upper bound $Re(\la_l) <<
\sqrt{N}$  on Bethe string roots was identified, beyond which significant
deformation of strings was
 reported \c{shastry00}.

However
 inspite of the above  major  criticisms of the
string hypothesis, strangely enough
  no significant  attention was given   in subsequent years 
 to understand or resolve this problem.
Nevertheless, as far as we know,
the  consistency of the  Bethe  string form itself  has never been
questioned before.
Therefore, it is  even more  surprising  for us  to find  that,  
  the well      known   
 Bethe string  for all the fundamental  models mentioned above, 
 widely quoted and used for so many years, is in fact not a precise  
 solution to the BAE even  
 within its validity range and for  large  $N$.
   After noticing this we are able  also  to obtain here
 the precise and  consistent form of the Bethe string by refining its
 correction terms, which  therefore  should  replace   
the accepted 
 string forms  presented in a series of celebrated works
\c{bethe31,taka71a,gaudin71,taka71b,lai71,taka72,kor94,schlot87,foers93} 
 dealing with the fundamental spin and electron models
cited above.
Moreover  we have   discovered
  a new non-Bethe string with arbitrary $r$ roots,
  compatible with the earlier observations. Based on 
 the  explicit forms of two kinds of strings
 one could  compare therefore their singularity structures
appearing in the BAE, the energies   of the
 states formed by them 
and hopefully 
all other  thermodynamic quantities, which  
should  contribute toward the solution of 
 the  longstanding problem of  string hypothesis.

Let us start by checking  first the known Bethe string
as solutions to the respective BAE at the thermodynamic limit.
 In   all the  
  spin and electron 
 models mentioned above,
  the same BAE are obtained by both   coordinate and the algebraic 
 Bethe ans\"atze, though  in the former  they are generated 
as a consequence of the periodic boundary condition on the wave
functions \c{bethe31}, 
while in the later 
 as the analyticity  condition for  
 eigenvalues of the transfer matrix \c{TakhFad81}.  
  $XXX$ spin-$\ha$ chain being a generic model,
we intend to examine its solutions in more detail and conclude  about 
the other models by analogy. 
\\ \\
{\it $XXX$-spin chain}:
The BAE for this model may be given as
\be V_{\ha}^N(\la_l)=\prod_{n\neq l}V_{1}(\la_l-\la_n), \qquad
l=1,\ldots,r,
\ \  
V_{\alpha}(\la)\equiv 
  {\la+i \al \over \la-i\al}. 
\ll{bel}\ee
The well known 
 Bethe string (BS) solution to the above BAE, valid for large $N$, is usually    
  given by
\be 
 \la^{(bs)}_l=\la_0 +i \eta (l-{r+1 \ov 2}) +i O
(e^{-\alpha N})
, \qquad  l=1,\ldots,r, \ \ \al      >0,
\ll{bs}\ee
with $\eta=1$.
This form  of   solution
 suggests that, exponentially small
 corrections: $O(e^{-\alpha N}) $
are needed  for
    large values of $N$, which presumably
 are of the same order  due to the same $\alpha  $ parameter appearing
  for all the roots.
%
%
At $N \to \infty$ therefore  all corrections 
in (\re{bs})  must vanish making it  an apparently   
 exact solution \c{TakhFad81,Izum88}.
Note that conjugate roots are included  in  (\re{bs}) with  
$ \la_l^* = \la_{r+1-l}, ~~  l=1,\ldots, s, $
 where $s= {r \ov 2}$ for $r$ even and
$s={r-1 \ov 2}$ for $r$ odd.

To follow the  idea of this solution we note that,
  any complex solution with $Im(\la) <0 $
 always makes 
 $|V_{\ha}(\la)|< 1$, which can   be proved for 
the Bethe string with    $l \leq s $ by inserting
the explicit form (\re{bs}) with  arbitrary $r$ : 
   \be
~|V_{\ha}(\la^{(bs)}_l)|=
|\left( {\la_0 +i (l-{r\ov 2})
 \over \la_0+i (l-{r+2 \ov 2}) }\right)| =(1+\kappa_l)^{-\ha} \equiv 
e^{-v_l } <1, \ll{kappa}
\ee
due to $ \kappa_l={r+1-2l \ov
\la_0^2+({r \ov 2}-l)^2} > 0$.
 Consequently, for large values of $N$
the LHS of BAE  (\re{bel}): 
   $~V_{\ha}^N(\la_l^{(bs)}) \sim O(e^{-v_l N})$
 must always vanish exponentially. Therefore for the  Bethe string to be a
consistent 
solution,  the RHS must also contain an exponentially vanishing factor of the
same order of smallness. 
By direct check one  sees easily that 
in the simplest cases of $r=2,3$ this condition is fulfilled and hence
, the Bethe string    satisfies
(\re{bel}) for large $N,$
as shown explicitly in \c{TakhFad81,Izum88,condmat00}.
 
However, though at this point it  might seem 
natural  to assume that,  
 the same argument  must go through for any arbitrary
$r$  \c{TakhFad81,Izum88}, we will see by direct insertion that this
  analogy fails here and
 the RHS  as such becomes
inconsistent    for
the string form (\re{bs}) starting from $r=4$.
One finds  that 
 among  the factors in RHS, one with the next higher root 
gives for the Bethe string (\re{bs}): 
$ V_{1}(\la^{(bs)}_{l}-\la^{(bs)}_{l+1})=V_{1}(-i +~O(e^{-\al N})) 
\sim ~O(e^{-\al N}),$ an exponentially small term.
However one can also  notice immediately that 
for $l >1$ and $r\geq 4$,
  another singular factor  appears from the adjacent lower
root yielding an exponentially large term having the same order:
$ V_{1}(\la_{l}^{(bs)}-\la_{l-1}^{(bs)})
=V_{1}(+i +~O(e^{-\al N})) \sim ~O(e^{\al N})$,
while the rest of the  factors coming from other roots  give only finite
contributions. Multiplying all these factors we finally get the RHS of
(\re{bel}) as $~O(e^{-\al N}) O(e^{\al N}) \sim O(1)$, which however
  has
a finite  limit 
contradicting the vanishing LHS.
 Therefore we see that for all roots $\la^{(bs)}_l, $ with
$l=2,\ldots, r-1 $ and $ \ r \geq 4,$
 having  two adjacent neighbors $\la^{(bs)}_{l\pm 1},$
 the  well known string  form 
(\re{bs}), as such,
 is not a consistent solution of BAE
(\re{bel}), except  only for the  end-roots $l=1,r$
 and  the real root, which cover also the cases $r=2,3$.
 For example 
 for $r=4$, as worked out in our  preliminary report 
 \c{condmat00},  the above string  form holds 
 for the end-roots with $l=1,4$, but not   for the roots with $l=2,3$.
\\ \\
{\it $XXZ$-spin chain}: The BAE   take  
 the same form as in $XXX$:
 $\ti V_{{\eta \ov 2}}^N(\la_l)=\prod_{n\neq l}\ti 
V_{\eta}(\la_l-\la_n),$  but through a  redefined function 
$\ti  V_{\alpha}(\la)\equiv 
  {\sin ({\la+i \al}) \over \sin ( \la-i\al) }$ and its well known string
solution for $\la^{(bs)}_l$
 can be given again by (\re{bs}) 
 \c{taka71a}. Skipping the details we mention only that, since the
functions $\sinh x$ and $x$ behave similarly at small 
$x$, the above reasoning for the $XXX$-string  goes parallelly in this case
 and one encounters a similar missmatch  for its  standard string form. 
\\ \\
{\it Repulsive $\delta$-function electron gas}:
For this and all other electron models considered below an additional set of
rapidity variables is needed.
The BAE is therefore extended to include another set of equations
$\prod_j^{N_e} V_{{c \ov 2}}(\la_l-k_j)=\prod_{n\neq l} 
V_{c}(\la_l-\la_n),$ though its structure is very similar to (\re{bel})
and the string solution for $\la_l$ is given in the same form (\re{bs}) with
$\eta=c >0$ and  $\{k_j\}$ real \c{taka71b,lai71}.
 Since here $k_j$'s are real, for  complex $\la_l^{(bs)}$ with $l \leq s,$ 
the LHS    becomes a product of terms each being $<1$.
Therefore for the string solutions at large $N_e,$
 one gets a vanishing  LHS, while the
RHS having exactly the same form as in (\re{bel})
 remains finite as argued above.
\\ \\
{\it Hubbard model}:~~
The additional BAE exhibit very similar structure to the above electron 
model giving $~~\prod_j^{N_e} V_{{U \ov 4}}(\la_l-\sin k_j)=
\prod_{n\neq l}
V_{{U \ov 2}}(\la_l-\la_n).$ However in this case along with the string
solutions for $\{\la_l, \la'_n\}$ in the form (\re{bs}) with $\eta={U \ov
2}$, $M'$-pairs of solutions from  $\{k_j\}$ can also be of 
 string type  satisfying
$k_n^\pm-\la'_n= \mp i{U \ov 4}, \ n =1, \ldots, M'$  \c{taka72}. Note that
inspite of $k_n^\pm$ being complex, due to the presence of 
 their conjugates  also 
in the factors of
 LHS, any complex string root for $\la_l,\ \ l \leq s$ will make 
the factors $<1$. Therefore for $N_e \to \infty $ the LHS $\to 0$, while
 the RHS
 due to its same form as in the above discussed cases, gives
finite contribution for $r \geq 4$.
\\ \\
{\it Supersymmetric $t-J$  model}:  For  a particular
(BBF) type of  excitations   \c{foers93}, one  set of BAE  takes
the form
$V_1^N(\la_l)\prod_{\beta}
V_{{ 1}}(\la_l-\gamma_\beta)= \prod_{n\neq l} V_{{ 2}}(\la_l-\la_n).$ The
string solutions for $\la_l$ are 
given again by   (\re{bs}) with all $\gamma_\beta $'s real \c{foers93}. 
Noticing  the first factor in the LHS to be the  same as in 
$XXX$ case and  $\{\gamma_\beta\}$ being real, we conclude as before
 that LHS $\to 0 $ for $N \to \infty.$ 
 The RHS however being
 same again as in $XXX$ gives  non-vanishing  terms.
starting from $r\geq 4$.

As we see from the above arguments, the 
missmatch of the well known Bethe 
string form (\re{bs}) is due to its imprecise
correction terms. We therefore    
propose the consistent and precise form of the Bethe string (PBS)  
 as
\be 
 \la^{(pbs)}_l=\la_0 +i (l-{r+1 \ov 2}) +i O(e^{-\alpha_l N})
, ~~~  l=1,\ldots, s ~ \  \ \mbox {and} \  \ \la_{r+1-l}=\la_l^*,  
\ll{bsr}\ee
provided
the exponential orders in its correction terms 
 are fine-tuned as a strictly growing sequence 
\be
0<\alpha_1 <  \cdots <\alpha_l<\alpha_{l+1}<\cdots<\alpha_s,
\ \mbox{with}~ \alpha_l-\alpha_{l-1}= v_l=\ha ~ln
(1+\kappa_l) >0,
\ll{grow}\ee
with $ \kappa_l$ as defined in (\re{kappa}).
Note that the term $O(e^{-\alpha_l N}) $ in (\re{bsr}) stands for the terms
like
$c_l e^{-\alpha_l N} $ with $l$-dependent multiplicative constants $c_l .$ 
The essential
point in proving  the validity of this refined Bethe string   is that
unlike the known form (\re{bs}) 
the adjacent roots contribute now zeros and poles of different orders of
smallness in the RHS of BAE, since 
$ V_{1}(\la_{l}^{(pbs)}-\la_{l+1})^{(pbs)}=
O(e^{-\al_l N})-O(e^{-\al_{l+1} N})\approx 
O(e^{-\al_l N}),  $ while $ V_{1}(\la_{l}^{(pbs)}-\la_{l-1}^{(pbs)})=
\left(O(e^{-\al_l N})-O(e^{-\al_{l-1} N})\right)^{-1}\approx 
O(e^{\al_{l-1} N})$, using the strict inequality  (\re{grow}).   Therefore
 the RHS becomes $ O(e^{-\al_l N})O(e^{\al_{l-1} N}) \sim O(e^{-v_{l} N})$,
 i.e. consistent with the LHS, which has  the same vanishing limit as before.
 It is important to note that,
 contrary to its known form 
  (\re{bs}), the correction terms present    in  
the proper  Bethe string (\re{bsr}) are
 rather complicated with simultaneous involvement 
of  small terms of all different   orders and 
  none of them can  be neglected from the beginning,
  even  at $N \to \infty $.
%

We emphasize again  that, $XXX$-spin chain being a generic case, the same
 consistent Bethe string (\re{bsr}) with refinement (\re{grow}), will be
 equally valid for all fundamental integrable models discussed above 
and should therefore
 replace the corresponding well known and widely used 
Bethe strings appearing in
 related works cited above.
 Curiously however this seems not to affect
  string based physical results
obtained through   TBA. The TBA method \c{taka71a,taka71b} 
  needs  apparently 
not  the  solution of individual   BAE, but   
 the solution of a   product of several  BAEs with different roots 
for the same string: $ \prod_\al V_{\ha}^N(\la_{\al}^{(1)(bs)} )=\prod_{\al,
 \j}V_{1}(\la_\al^{(1)(bs)}-\la_j^{(2)(bs)})$.
Note that  since  the complex conjugate of each root is  also present in
the product,
 its LHS remains finite even for large $N$ 
and therefore the standard string solutions hold for such 
product-BAE. 
    Therefore inspite of
 being imprecise in form for  the individual BAE, the known Bethe
  string fortunately is capable of
 producing correct TBA results.

 Nevertheless 
the precise form of the Bethe string we find here is important not only as a
 consistent form of solution for the BAE but also for
 its possible comparison with a non-Bethe solution. 
 For comparison   one
 needs  also      an explicit non-Bethe string 
   with arbitrary $r$ roots, which however is  
 not   available
in the literature.
We  therefore propose as well a new  non-Bethe
string (NBS) solution to    BAE   (\re{bel})  in the form
\be 
 \la^{(nbs)}_l=\la_0 +i {1\ov s} (l-{r+1 \ov 2}) +i O
(e^{-\alpha N}),  \ ~~ ~  l=1,\ldots, r, \ \ \al >0,  
\ll{as}\ee
 with $s={r \ov
2}$ for even $r$ and   $s={r-1 \ov
2}$ for  odd $r$.
Note that the correction terms in (\re{as}) are much simpler  
compared to  the Bethe case and
moreover they can  be regulated by the background real roots,
the presence of which is essential, as we see below, for the survival of the
non-Bethe strings.
For simplicity we consider  only the case of even $r$; the odd $r$ case 
 can  be handled in a similar way. It is easy to check that 
 each root
$\la_l^{(nbs)}$ with $l \leq s$
for  string 
 (\re{as}) yields as before LHS $=O(e^{- v_l N}),$ now
with $ v_l=\ha \ln (1+\kappa_l), \ \kappa_l={ 1 \ov D}(r+1-2l), D={r \ov
2}(\la_0^2+{1 \ov r^2}(2l-1-{r \ov 2})^2)
$, while  the RHS 
 having   only one partner $ \ti \la_l^{(nbs)}= \la_{l+s}^{(nbs)}$
contributes  also with  a vanishing term of the  order:
 $ V_{1}(\la_{l}^{(nbs)}- \ti
\la_{l}^{(nbs)}) = O(e^{-\alpha N})$.
However this apparent consistency 
breaks down if we consider the product of this 
 equation  with that of its partner giving in the LHS
again an exponentially  vanishing term 
$ |V_{\ha}(\la_{l}^{(nbs)})V_{\ha}(\ti \la_{l}^{(nbs)}|^N  \approx
O(e^{-( v_l - v_{l+s}) N})$ for large $N$ and 
$l < {1\ov 4}(2+r)\equiv l^*.$ For $l >  l^*$  this term 
 similarly blows up 
exponentially. 
 However   one finds now  that the singular terms in the
corresponding  RHS get
 mutually canceled due to opposite but equal contributions from 
$l$-th and its partner, i.e. $ (l+s)$-th root, which    leads therefore the RHS
 to a finite limit
 and thus to a  contradiction.
Therefore we conclude that  unlike the Bethe case such non-Bethe strings
 can not survive in
isolation and  must be coupled with background real roots, which
can make  ansatz  (\re{as}) 
consistent by  supplying
 necessary vanishing (exploding) terms in the RHS.
This also shows perhaps at the microscopic level why the non-Bethe strings
were observed earlier only for the aniferromagnetic ground state (AFGS) \c{Woy82}. 
For demonstrating  this  we couple  (\re{as}) with $M$ number of real roots
$\{\nu_a\}$ and notice that  due to the appearance of additional
terms like $  \prod_a^M|V_{1}(\la_{l}^{(nbs)}-\nu_a)|$   in the RHS  for 
the  $l$-th root of the string,  a new  exponentially diminishing (increasing)
 term  may arise and and since each of its constituent terms
are either $<1$ or $>1$,  by multiplying them one gets  the contribution
 from
$M$ number of such  terms  as
$ O(e^{-\sum_a^M\phi_{la}})$, where $\phi_{la}
=\ha \ln (1+\kappa_{la}),$ with $ \ \kappa_{la}={1  \ov \bar D}(r+1-2l),
\bar D={r \ov
4}((\la_0-\nu_a)^2+{1 \ov r^2}(2l-1)^2)
$. 
Therefore for  large $M$  there appears now a   possibility   to
compensate for the singular term arising in the LHS of the product BAE
mentioned above and  thus
to resolve  the  controversy for the existence of strings like 
 (\re{as}).
 The exact values of $\nu_a$ should in fact be determined
selfconsistently from the BAE. However this is difficult to achieve 
in practice  for large
values of $M, N$, even numerically and such strings can exist
only for large $N, M$. Therefore  we  make here only some 
 estimates to show the
right trend. Since the product of the BAE's for the 
$l$-th root with its partner leading to a contradiction for the
isolated NBS is our main concern,  we observe  first that 
each of the additional terms in its RHS, e.g. $
e^{-(\phi_{la} -\phi_{(l+s)a})}$  exhibits 
 diminishing   (increasing) trend also
for $l<l* (l>l*)$,  matching exactly with the behaviour  of the LHS
established above.
 This shows at least qualitatively that for the NBS
 the coupling with  real
 roots can indeed yield necessary singular terms in the RHS 
 that might match
  such terms appearing in the
LHS.
For having  further  quantitative estimates we have checked 
numerically the ratios of the singular terms arising in both the sides
and  sought for the real solutions of $\nu_a$ when they match.
 We
 find in particular that for various values of $r$ and $l$ there  always exists a pair of
real solutions for  $\nu_a$
to the relation:  $e^{-2( v_l - v_{l+s})}
\approx e^{-(\phi_{la} -\phi_{(l+s)a})}$
 taking one positive and another negative values. For 
$\la_0=0$ for example, we get   $\nu_a$ placed symmetrically around the
origin, which  gives  the total momentum of the system:$ P=2 \pi\cdot
Integer $, in agreement with the translational invariance of the system.
Therefore for $M={N \ov 2}$ and 
a   distribution of $\nu_a$'s having the same order  in magnitude
   one is likely to match the singularities in both sides of the BAE 
 allowing the NBS (\re{as}) to exist.
Note that in the thermodynamic limit this distribution of real roots should
correspond to  
the AFGS and the NBS should correspond to
 the earlier non-Bethe string 
results observed also over the AFGS
\c{DesLow82,Woy82,BabdeVVia83,Woy83}.
Now we focus on to some interesting properties of NBS 
(\re{as}) and show its more resemblance 
with earlier observations.
Firstly we notice 
  that due to the reduced inter-root distance $\Delta={i \ov s}$ its 
 length  $L(r)= i{({r-1 }) \ov s}$ is bounded as  $ 1\geq |L(r)| \geq 2$
and   in contrast to  the Bethe string 
 exhibits therefore  a { close-root} form, which is in accordance with the
earlier observations.
It is also evident that together with each root ($\la_l$)
 its partner ($\ti \la_l$), its conjugate ($ \la_l^*$) and
 the partner of its conjugate ($\ti \la_l^*$) form a closed
unit
  of four  and being decisive contributors to the equation at large $N$,
 become
  almost independent entries at the thermodynamic limit. These groups of
four may also get reduced to form a { doublet}
 or a { triplet}
due to possible degeneracies (see fig.1). However this can occur only once
and that also when
$r$ is not divisible by $4$. 
  Thus the  close-root string (\re{as})
 may split up into units 
of four, three and two at  the thermodynamic limit, reproducing again the
non-Bethe structures observed earlier. 
Moreover, we see that for each of the roots with $l<l*$, which corresponds to
$|Im (\la_l)|> \ha$, its partner  $\ti \la_l$ is always with 
 $|Im (\ti \la_l)|< \ha$. This fact  also mimics amazingly the observation
of  \c{BabdeVVia83} stating that the close root of type I 
must have  their partners from  among
the  close roots of type II. Due to such striking agreement with 
earlier studies  we hope that the more general non-Bethe 
 string structure  (\re{as})   should survive also 
in the thermodynamic limit and should be consistent with the integral
equations derived earlier for the complex roots in the anti-ferromagnetic
case.
The precise and explicit  forms of the general 
  Bethe (PBS) (\re{bsr}) and non-Bethe
(NBS)
(\re{as}) $r$
strings  found here should also  be useful for comparing their corresponding
properties.
As we have noticed,
their correction terms  
 induce different nature of singularity
 structures in the BAE at large $N$.
Generically   roots of the Bethe-string (PBS) (\re{bsr})
  produce  zeros and poles  of different
 orders of smallness in the BAE,  arranged in a growing sequence
 and  the roots  are located mostly
 in the wide-root region:
 $|Im(\la_l)|> 1$.
The non-Bethe string (NBS) (\re{as}) on the other hand,
allows  the singularities of its roots  to
 be of the same order
of smallness and the roots themselves are 
concentrated only in the close-root region.
 Therefore,
 if one looks for the complex roots of the BAE 
at the AFGS  with   
 their singularities having the 
same order in the whole
 complex plane,
   then among the string solutions  only the  close-root NBS
  are likely to appear, which supports the earlier findings.
 Similarly the  energies of the states created by Bethe and
non-Bethe $r$-strings can also be compared  
using their explicit forms giving
$ E^{(bs)}_r
={r\ov \la_0^2+({r \ov 2})^2}$
and $~
E^{(nbs)}_r=
\sum_{j=1}^s {2 g_j \ov \la_0^2+g_j^2},
~$ respectively,
where   the factor  $g_j=g_1-{1 \ov s}(j-1)$, with $g_1={3 \ov 2}$
 for odd  and
 $g_1={3 \ov 2}-{1\ov 2s}$ for even $r$.      
One can show   analytically for $r \to \infty$ 
and numerically for finite $r$    (see fig. 2)
that  $ E^{(bs)}_r$  is always lower than    
   $ E^{(nbs)}_r$, while both 
    have lower values than  that of the $r$-free magnons: $
E^{(free)}_r= 
{r\ov \la_0^2+{1 \ov 4}}$.
 Therefore one   concludes    
  that, though both Bethe and non-Bethe strings may give bound states,
  the  non-Bethe ones must be  
  more loosely bound with higher energies.
However at this stage conclusive statements
are still difficult to make regarding    the string hypothesis problem. 
One perhaps should  consider  the  TBA analysis  using  
  the  general $r$-NBS over the AFGS, which we leave as a future problem.   
\\ \\
Acknowledgment: I would like to thank Prof.  Deepak Dhar and Prof. Andreas
Kl\"umper
 for stimulating discussions. I also acknowledge with thanks
 the hospitality 
of the Institute for Theoretical Physics,   Dortmund
University, where this work was completed and  the
financial and other supports from the AVH Foundation, Germany.

\newpage
\begin{figure}[ht]
\begin{center}
\leavevmode
\epsfxsize=0.65\textwidth
\epsfbox{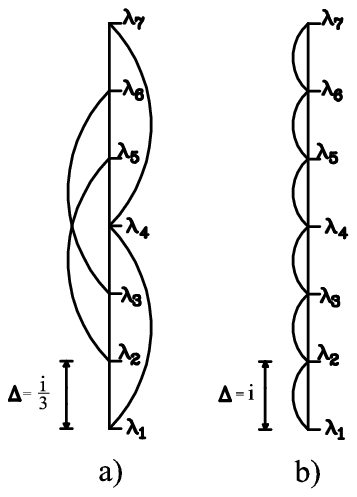}
\end{center}

\vskip .8cm

\caption{
Arrangement of roots in string solutions with $r=7$.
a) Non-Bethe string with length $L(7)=6 \Delta= 2i$ represents a close-root
form and breaks up into a quartet with $ \{ \la_2, \ti \la_2(= \la_5),
 \la_2^*(=\la_6) ,
\ti \la_2^*(= \la_3) \}$ and a triplet with
$ \{ \la_1, \ti \la_1(= \la_4)=\ti \la_1^*,
 \la_1^*(=\la_7)  \}$.
b)  Bethe string  with length $L(7)=6 \Delta= 6i$ in the wide-root region
can not break up into smaller units since each root has two partners, each
of which  in turn has a different partner. 
}
\label{fig. 1}
\end{figure}
\begin{figure}[ht]
\begin{center}
\leavevmode
\epsfxsize=0.65\textwidth
\epsfbox{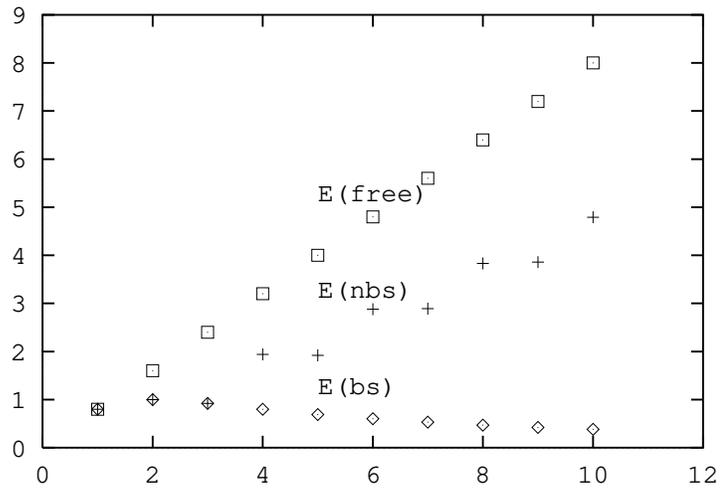}
\end{center}
\medskip

\vskip .5cm

\caption{
Comparison of energies  corresponding to 
$r$-Bethe string: $E(bs)$, $r$-non-Bethe string: $E(nbs)$ and $ r$ free
magnons: $E(free)$ for $r=[1:10]$, showing both strings as bound states with
$ E(free) \geq E(nbs) \geq E(bs) $.
}
\label{fig. 2}
\end{figure}

\end{document}